\documentstyle[prl,aps]{revtex}
\begin{document}
\draft  
 
\title{
X-ray Absorption Fine Structure in Embedded Atoms  }

\author{J. J. Rehr,$^{(1)}$ C. H. Booth,$^{(2)}$  F. Bridges,$^{(2)}$
S. I. Zabinsky$^{(1)}$}

\address{$^{(1)}$Department of Physics, University of
Washington, Seattle, WA 98195}

\address{$^{(2)}$Department of Physics, University of California Santa
Cruz, Santa Cruz, CA 95064}

\date{\today}
 
\maketitle

\begin{abstract}
   Oscillatory structure is found in the atomic background absorption 
in x-ray-absorption fine structure (XAFS).  This atomic-XAFS or AXAFS arises
from scattering within an embedded atom, and is analogous to the
Ramsauer-Townsend effect.  Calculations and measurements 
confirm the existence of AXAFS and show that it can dominate 
contributions such as multi-electron excitations.
The structure is sensitive to chemical effects and thus
provides a new probe of bonding and exchange effects on the scattering
potential.

\end{abstract}

\narrowtext
\twocolumn

    The main features of X-ray absorption spectra $\mu$(E)
are due to one-electron transitions from deep core levels.
In molecules and solids, oscillatory fine structure exists in $\mu(E)$
due to scattering of the photoelectron by neighboring atoms.  The
well known technique of x-ray-absorption fine structure (XAFS), which
includes both extended-XAFS (EXAFS) and x-ray-absorption near-edge
structure (XANES), is based on the analysis of this
fine structure.  In XAFS the oscillatory part
$\chi$ is defined relative to an assumed smooth ``atomic background''
absorption $\mu_0(E)$, i.e., $\chi=(\mu-\mu_0)/\mu_0.$
   A complication is that $\mu_0(E)$
is not necessarily smooth.  For example, the background may exhibit
such well known structures as white lines, resonances
and  jumps due to multi-electron
transitions, even well above threshold.
  Less well known, however, is the possible fine structure in 
$\mu_0(E)$ itself, in molecules and condensed
systems, as discussed by Holland {\it et al.}\cite{Holland78}
The purpose of this Letter is to show that this atomic
x-ray-absorption fine structure (AXAFS) can produce large oscillations,
has an  XAFS like interpretation, and can alter XAFS analysis.
In view of recent advances in XAFS theory and analysis
techniques\cite{FEFF5X,Li92,Newville93}, in which the background
plays a crucial role, this structure is now particularly important.

This extra fine structure originates from resonant scattering
``in the periphery of the absorbing atom"\cite{Holland78}.
The effect is  like  an internal Ramsauer-Townsend (RT)
resonance where the incident electron is a spherical wave created at
the center of the atom, rather than a wave scattered by an atom.
As the photoelectron electron approaches a potential barrier - in this case the 
edge of an embedded atom potential -
the reflection coefficient oscillates with energy, with a
pronounced increase just above threshold,  followed by a dip and subsequent
oscillations that conserve integrated oscillator strength.
We find that AXAFS can be the dominant
background fine structure and has features in the same energy range
as multi-electron transitions, complicating detection of the latter.
Using a new background subtraction technique\cite{Li92},
 experimental backgrounds for Ba, Ce and Pr K-edges
are obtained which exhibit AXAFS as large as 60\% of the XAFS
amplitude.  Theoretical calculations based on an 
{\it ab initio} XAFS/XANES code FEFF 5X \cite{FEFF5X}
confirm these observations. To our knowledge, the only previous
attempt to identify AXAFS\cite{Holland78} was only partly successful.
Notable discrepancies between theory and experiment were found at low
energies and the work did not derive its oscillatory character. 
We believe, however, that evidence for AXAFS exists in many previous studies,
although not heretofore identified as such.  In particular we suggest that
AXAFS is largely responsible for the spurious peak at about half the first
neighbor distance often observed in XAFS Fourier transforms\cite{Frenkel93,Li92}.



   A number of improved background subtraction techniques have recently been
developed\cite{Li92,Newville93}. The approach adopted here\cite{Li92}
is based on an iterative procedure, which is summarized as follows:
After removing the pre-edge absorption, a smooth spline function is fit to
the absorption data $\mu(E)$ (Fig. \ref{fig1}), which simulates to lowest order the free
atomic absorption $\mu_{0}(E)$ without XAFS or other features present.
Then a trial XAFS function $\chi(E)=[\mu(E)/\mu_0(E)]-1$ is obtained.
A Fourier transform of $\chi$ with respect to wave number $k$ defined
with respect to threshold energy $E_0$,
yields peaks in $r$-space corresponding to the distribution of neighbors
to the absorbing atom.  Initially, the
transforms often have a spurious, $r$-space peak near 1 \AA\,
that is inconsistent with their known structures.
Next an approximate fit of the first few peaks of the $r$-space
transform is made using theoretical XAFS standards\cite{FEFF5}.
This fit is transferred back to energy space and subtracted from the
experimental data to remove most of the low frequency XAFS oscillations.
A high order spline is used to smooth the remaining data (Fig.\ \ref{fig1}).
The positions of the knots in this spline are varied
to follow the larger features in the residue. 
This spline function then becomes a new $\mu_{0}(E)$ and a new XAFS function
is extracted.
The process is iterated to convergence, typically in several iterations. 
With this procedure the background $\mu_0(E)$ contains all the atomic
fine structure and the spurious $r$-space peak near 1 {\AA} 
is eliminated.

This procedure was tested on a theoretical absorption spectrum from FEFF 5X
for a model of PrBa$_{2}$Cu$_{3}$O$_{7}$ (PBCO) which included many XAFS
shells and a background with the above RT-like resonance.
PBCO was chosen because the contribution from higher shells is
significant. This check therefore tests both the fit
to the XAFS and to the background.
The extracted background fits the simulated background well (Fig.\ \ref{fig1}).

 This procedure was then applied to Ba, Ce, and Pr K-edge data 
of BaO, CeO$_{2}$, and PBCO.  The data were collected at $T\simeq$ 80 K,
at the Stanford Synchrotron Radiation Laboratory (SSRL) using (400) monochromator 
crystals.  Details are given in a separate paper\cite{Booth93}. 
The extracted backgrounds of all these high energy K-edges (Fig.\ \ref{fig2})
are similar in shape and energy scale, exhibiting the near-edge peak and
dip structure consistent with that expected for a RT resonance.
The magnitude of this structure is comparable to EXAFS amplitudes and is
a factor of four larger than the step-like structures
observed above the edge for the rare gas Kr or for the Rb and Br K-edge data  for RbBr. 


 We now briefly discuss the theory of AXAFS and show that it
has an interpretation analogous to
the curved-wave theory of XAFS\cite{Rehralbers}.
``Embedded'' atoms in solids may be
defined in terms of their respective scattering potentials.
The final state potential $v_0$ at the absorption site consists of
a bare atomic potential $v_a$, plus extra-atomic contributions $v_e$ 
from the tails of the electron distributions of neighboring atoms. In
the muffin-tin approximation,
\begin{eqnarray}
 v_{0}(r)  =     v_a(r&) + v_e(r), \qquad & (r \leq R_{mt}), \nonumber \\
           =   v_{mt},&          \qquad & (r \geq R_{mt}), 
\label{eq1}
\end{eqnarray}
where $R_{mt}$ is the muffin-tin radius.
  For simplicity we consider a one-electron calculation of photoabsorption by
an embedded atom using the Fermi Golden rule and the dipole approximation,
i.e.,
\begin{equation} 
\mu_0(E) = {4\pi^2}\alpha \omega
\sum_f |\langle c|\hat \epsilon \cdot \vec r| f \rangle|^2
          \delta(E-E_f),
\label{eq2} 
\end{equation}
where $\alpha\simeq 1/137$ is the fine structure constant, $\omega$ is
the X-ray energy (we use Hartree atomic units $e=m=\hbar=1$),
$E=\omega-E_c$ is the photoelectron energy, 
$\hat \epsilon$ is the x-ray polarization vector, and
the final states $|f\rangle=(1/r)R_0(r) Y_{lm}(\hat r)$
are calculated at energy $E_f = (1/2) k^2$ in the embedded atom
potential $v_0$.
The normalized radial wave functions $R_0(r)$  are obtained by
matching the regular solution of the radial $l$-wave Schr\"odinger equation
to the asymptotic form $R_0(r)=kr[j_l(kr) \cos\delta_l - n_l(kr)\sin\delta_l]$,
$(r\geq R_{mt})$, where $j_l$ and $n_l$ are spherical Bessel functions,
$\delta_l$ is the $l-$th partial wave's phase shift, and $l$ is fixed by
dipole selection rules.  This matching procedure is equivalent to a
calculation of the Jost function $F_l(E)$ which guarantees final state normalization,
as discussed by Holland {\it et
al.}\cite{Holland78} and by Newton\cite{Newton}. 
In particular Holland {\it et al.} show that the atomic cross-section can be
written as $\tilde\mu_0/|F_l|^2$, where $\tilde\mu_0$ is a reduced matrix element
which varies smoothly with energy. 
All of the calculations of AXAFS reported here are based on
an analogous matching procedure for the relativistic spinor wavefunctions used
in FEFF, without any of the simplifying approximations of the following
discussion.  Additional details will be given elsewhere\cite{FEFF6}.

The formal relations\cite{Newton} satisfied by the Jost function,
are very general and do not explicitly show
the oscillatory behavior of AXAFS.
Thus to illustrate its nature we present a highly simplified model 
based on first order perturbation theory with respect to the free
atom potential.  We will assume that the free atomic background
has negligible oscillatory structure; sample calculations with large
muffin-tin radii support this assumption.
Using the spectral representation of the embedded atom
Green's function, the final state sum in Eq.\ (\ref{eq2}) can be expressed as
$\Sigma_f|f\rangle \delta(E-E_f) \langle f| =
(-1/\pi)\,{\rm Im}\, G_0$ where $G_0=(E-H_0+i0^+)^{-1}$
is the embedded atom Green's function (operator) and $H_0$ the
embedded atom Hamiltonian.  To first order
in the perturbation  $\delta v=v_0(r)-v_a(r)$, 
$G_0$ is given by $G_0\simeq G_a+G_a\delta vG_a$, where 
$G_a$ is the free atomic Green's function. 
For deep core absorption, the core states are highly localized so we need
only evaluate $G_0$ in position space for
very small arguments $r$ and $r'$, where $\delta v$ is negligible.
The radial part of $G_a$\cite{Newton},
is given by $G_a(r,r')=(-1/k)R_a(r_<)R_a^{+}(r_>)$, where
$r_{>(<)}$ is the greater(lesser) of $r$ and $r'$, and
$R_a^{+}=S_a + i R_a$ is the outgoing part of the 
radial Schr\"odinger equation.
Combining these ingredients, one finds that $\mu(E)$ can be
factored as in conventional XAFS theory\cite{Leependry}, i.e., 
$\mu_0=\mu_a(1+\chi_e),$
where $\mu_a$ is given by Eq.\ (\ref{eq2}) calculated with
free atomic states $|f_a\rangle$ and the AXAFS $\chi_e$ is 
\begin{equation}
 \chi_e \simeq - {\rm Im}\,{1\over k}
   \int_0^{\infty} dr [R_a^{+}(kr)]^2 \delta v(r). 
\label{eq3}
\end{equation}
An analogy to the curved-wave XAFS formula\cite{Rehralbers} is 
obtained by recognizing that the perturbation arises from the periphery of
the atom where one may approximate $R_a^{+}$
by its asymptotic form, $R_a^{+}\simeq c_l(kr) \exp(ikr + i \delta_l^a)$.  Here
$c_l(kr)$ is the curved wave factor\cite{Rehralbers}
in the spherical Hankel function $h^{(+)}(kr)=c_l(kr)\exp(ikr)/kr$.
For simplicity we model the perturbation as 
 $\delta v(r)\simeq v_{mt}/[1+\exp(\zeta(R_{mt}-r))]$, where $\zeta$
characterizes the decay of the atomic potential tails near $R_{mt}$.
 The integral (\ref{eq3}) can then be expressed as
\begin{equation}
\chi_e=-{1\over kR_{mt}^2} |f_e| \sin (2kR_{mt}+2\delta_l^a+\Phi_e). 
\label{eq4}
\end{equation}
where $f_e=|f_e|\exp(i\Phi_e)$ is an effective curved-wave
scattering amplitude. With the above model the AXAFS is analogous to a
damped harmonic oscillator, $f_e\sim\exp(-2\pi k/\zeta)/k$.
For comparison to experiment, the Eq.\ (\ref{eq4}) should have
a few additional factors as in the usual XAFS
formula, namely an amplitude reduction factor $S_0^2$, 
a Debye-Waller factor, $\exp[-2\sigma^2(R_{mt}/R)^2 k^2]$,
and a mean-free path term, $\exp(-2R_{mt}/\lambda)$.


 AXAFS comparisons between the theoretical calculations and experimental
results presented here are in reasonable agreement with each other
(Fig.\ \ref{fig2}), especially for the simple oxides.
The discrepancy at the edge for BaO is not fully understood, but may
point to errors in FEFF's muffin-tin potential and energy reference.
The long range oscillatory structure in the calculations is likely due to a
small discontinuity in FEFF's muffin-tin potential at $R_{mt}$. 

 To check whether multi-electron excitations might also be present, 
we used the Z+1 model to estimate where the step for a
two-electron excitation would begin.  In this model excitation energies
correspond to the ionization energies of Z+1 atoms, and are
99 eV for Ba and 113 eV for Ce, as indicated by arrows in Fig.\ \ref{fig2}.
Small features in the background 
were previously attributed to multi-electron excitations
based on this model\cite{Li92,Newville93}. However, it is likely that part
of the observed structure can also be attributed to AXAFS.

We point out that our calculations of the atomic backgrounds 
shown in Fig.\ \ref{fig2} were all done with ground state exchange 
potentials.  We found that the usual  Hedin-Lundqvist (HL) self-energy
model used in FEFF\cite{FEFF5} gives too large an
oscillatory amplitude near threshold. This is an indication of the
sensitivity of the AXAFS to the exchange interaction. Evidently
improvements to FEFF's muffin-tin potentials are necessary,
and AXAFS may be useful in assessing various improvements.


It is well known that simple, monotonic approximations to the atomic background
are not sufficient to obtain accurate XAFS data, emphasizing the
importance of improved background removal methods\cite{Li92,Newville93}.
However,   the atomic background $\mu_0$ and the XAFS $\chi$ are tightly
linked by the definition $\chi=(\mu-\mu_0)/\mu_0$, so the backgrounds
obtained for theoretical and experimental standards may differ.
Thus an understanding of AXAFS is essential to obtain experimental backgrounds.
 This difference also affects XAFS analysis; if one tries to isolate a
Ce-O standard without taking its oscillatory background into account,
one cannot obtain a good fit to the first Pr-O peak in PBCO. 
The inclusion of extra-atomic contributions in the atomic
background may at first seem arbitrary. For example, the XAFS could
be defined with respect to the bare atomic background, which is 
independent of the environment. However such a definition is
problematical and inconsistent with multiple-scattering theory based
on independent scattering sites; also because the exchange interaction is
not additive, it is not possible to construct the
scattering potential by superposing free atomic potentials.

 For the materials discussed in this paper, 
the AXAFS is quite large, and is the dominant contribution
to the background fine structure, exceeding multi-electron
effects in magnitude. {\it Ab initio} calculations of the AXAFS
agree reasonably well with these observations and with the simplified
model introduced here.  The size and character of these background features, 
particularly their interference with the first coordination shell peak, indicate
that accurate fits to XAFS data must take them into account.
AXAFS is also interesting in its own right, because it depends critically
on the scattering potential in the outer part of the absorbing atom. 
Thus, it provides a new and useful probe of
chemical effects, the electron self-energy, core-hole effects,
and other contributions to the embedded atom potentials. 

  We thank J.B. Boyce, G.G. Li, P. L\=\i vi\c n\v s,
M. Newville, B. Ravel, E.A. Stern, and Y.\ Yacoby 
for comments and discussions concerning background removal methods and
T.\ Claeson for assistance in collecting the data.  One of us (JJR) also
thanks the Institut f\"ur Experimentalphysik of the Freie Universit\"at
Berlin for hospitality, where part of this work was completed.
SSRL is operated by the U.S Department of Energy, Division of Chemical Sciences,
and by the NIH, Biomedical Resource Technology Program, Division of Research
Resources. The data was collected under University PRT time. 
This work was supported in part by DOE 
Grant DE-FG06-ER45415 (JJR and SIZ), and by NSF Grant DMR-92-05204 (CHB
and FB).

\begin{figure}
\caption{Top curves: Ce K-edge absorption $\mu(E)$ (dotted) and $\mu_0(E)$
(solid) from CeO$_{2}$
vs. energy $E$ above the Ce K-edge (40441 eV).  All spectra in this paper have 
their step height normalized to unity, and shifted as displayed.  The oscillatory 
structure above the edge
(dotted) is the XAFS.  The solid curve is the experimentally obtained
``atomic background'' absorption $\mu_0(E)$ (see text).  Note the
sharp dip in this background at $\simeq$ 115 eV.  Middle curves:
residue function and fit for Ba K-edge data from BaO.
 The residue functions are
the difference between the data and the fit in $E$-space, {\it i.e.},
$\mu_{res}(E)=\mu(E)/(\chi_{fit}(E) + 1)$. 
Bottom curves: simulated background (solid) and 
extracted background (dotted) as a test of our 
extraction method.}
\label{fig1}
\end{figure}

\begin{figure}
\caption{Experimental (dotted lines) and theoretical (solid lines) background
absorptions, $\mu_0(E)$, for the Ba, Ce, and Pr K-edges of BaO, CeO$_{2}$,
and PBCO. $\Delta$E is the energy above threshold, {\it i.e.,} 37,444, 40,441,
and 41,991 eV for the Ba, Ce and Pr edges, respectively.  Both the experimental
and the theoretical backgrounds have been adjusted to fit the Victoreen formula 
with a 4th-order polynomial.
The calculations are currently
limited by discontinuities at $R_{mt}$ which can effect the
AXAFS amplitude.  The BaO calculation has 
an additional threshold energy shift of +20 eV. Arrows indicate the
positions of Z+1 excitation thresholds.}
\label{fig2}
\end{figure}

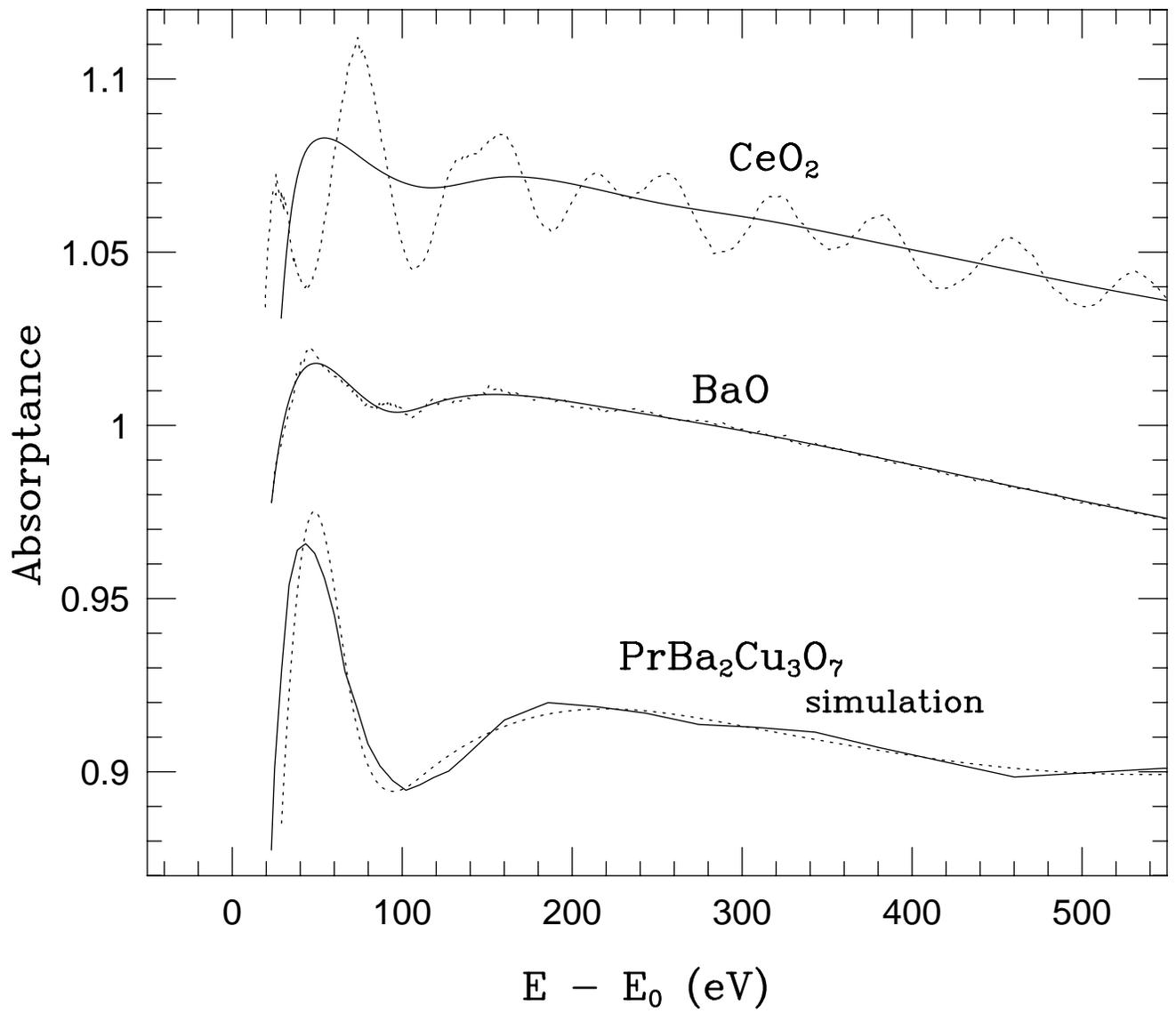
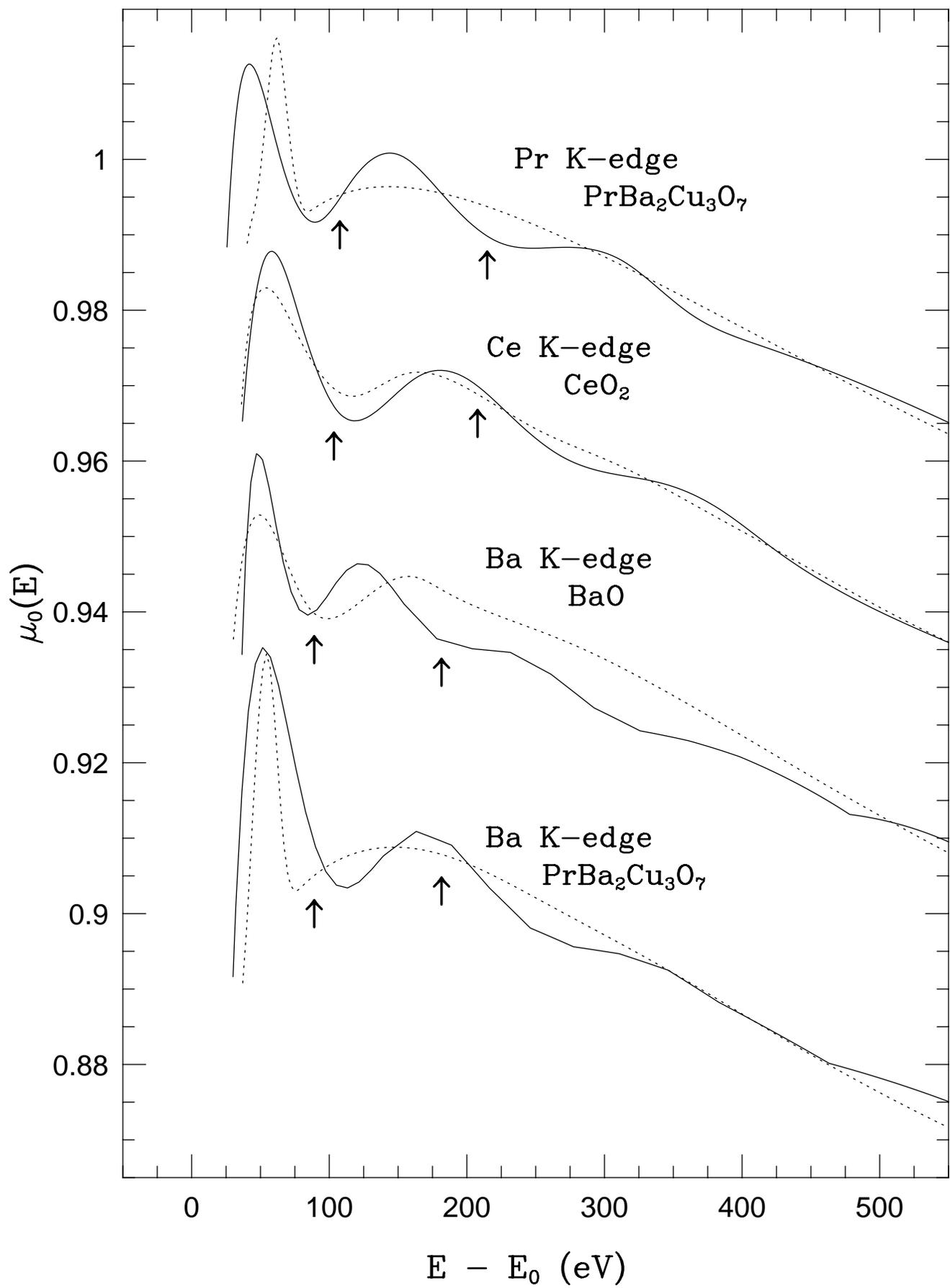
\end{document}